\documentclass[aps,prl,twocolumn,groupedaddress,showpacs,floats]{revtex4}
\usepackage{graphics}
\usepackage{amssymb}
\begin{document}

\title{Radiation-Induced Magnetoresistance Oscillations in a 2D
Electron Gas}
\author{Adam C. Durst, Subir Sachdev, N. Read, and S. M. Girvin}
\affiliation{Department of Physics, Yale University, P.O. Box 208120,
New Haven, CT 06520-8120}
\date{January 29, 2003}

\begin{abstract}
Recent measurements of a 2D electron gas subjected to microwave radiation
reveal a magnetoresistance with an oscillatory dependence on the
ratio of radiation frequency to cyclotron frequency.
We perform a diagrammatic calculation and find radiation-induced
resistivity oscillations with the correct period and phase.
Results are explained via a simple picture of current induced by
photo-excited disorder-scattered electrons.  The oscillations
increase with radiation intensity, easily exceeding the dark resistivity
and resulting in negative-resistivity minima.  At high intensity, we identify
additional features, likely due to multi-photon processes, which have yet to be
observed experimentally.
\end{abstract}


\maketitle

The electrical transport properties of a 2D electron gas (2DEG)
in a perpendicular magnetic field have been studied extensively over the past
two decades in connection with the quantum Hall effects.  However, recent
experiments, in which such systems are subjected to electromagnetic radiation,
reveal a surprising new phenomenon.  The initial experiments of
Zudov {\it et al.\/} \cite{zud01}, as well as more detailed subsequent studies
conducted by Mani {\it et al.\/} \cite{man02} and Zudov {\it et al.\/} \cite{zud02},
show that a peculiar oscillation of the longitudinal resistance is induced
by the presence of microwave radiation in systems at high filling factor.
Unlike the familiar Shubnikov-de Haas (SdH) oscillations which are controlled by the
ratio of the chemical potential, $\mu$, to the cyclotron frequency, $\omega_{c}$,
these radiation-induced oscillations are controlled by the ratio of
the radiation frequency, $\omega$, to the cyclotron frequency.
According to Ref.~\onlinecite{man02}, the minimum resistance values are
obtained near $\omega/\omega_{c} = \mbox{integer} + 1/4$.  These results are
surprising since naively one would only expect a peak at $\omega=\omega_{c}$
due to heating at the cyclotron resonance.  Interest in this phenomenon was
heightened when, using high mobility samples, both
Mani {\it et al.\/} \cite{man02} and Zudov {\it et al.\/} \cite{zud02}
observed that as the radiation intensity is increased, the minimum resistance
values approach zero and give rise to new zero-resistance states.  In this
Letter, we present calculations which reveal the origin of the resistance
oscillations, explain their period and phase, hint at the manner in
which zero-resistance states may develop, and predict higher-order effects,
likely due to multi-photon processes, which set in at high intensity.

\begin{figure}
\centerline{\resizebox{3.0in}{!}{\includegraphics{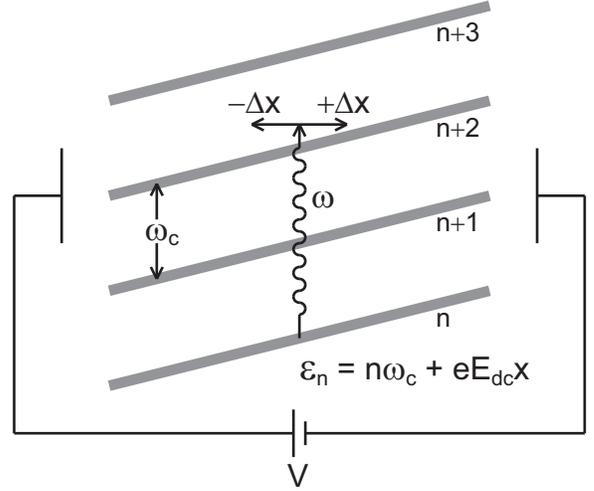}}}
\caption{Simple picture of radiation-induced disorder-assisted current.
Landau levels are tilted by the applied dc bias.  Electrons absorb photons
and are excited by energy $\omega$.  Photo-excited electrons are scattered
by disorder and kicked to the right or to the left by a distance $\pm \Delta x$.
If the final density of states to the left exceeds that to the right,
dc current is enhanced.  If vice versa, dc current is diminished.  Note that
electrons initially near the center of a Landau level (where the initial
density of states is greatest) will tend to flow uphill for
$\omega/\omega_c \approx \mbox{integer}+1/4$.}
\label{fig:toypicture}
\end{figure}

It is instructive to begin by exploring a crude treatment through
which a simple physical picture can be
established. The electronic states of 2D electrons without
disorder in a perpendicular magnetic field, $B$, are the Landau
levels, $\epsilon_{n}=n\omega_{c}$, where $\omega_{c}=eB/m^{*}c$,
the $n$ are nonnegative integers, and $m^{*}$ is the effective
mass. (Here, and always, we
set $\hbar=1$ and drop the constant energy shift
of $\omega_{c}/2$.) Each level has a degeneracy per unit area of
$1/2\pi\ell^{2}$ where $\ell=\sqrt{c/eB}$ is the magnetic length.
In the presence of disorder, the Landau levels are broadened. In
correspondence with experiment, we consider states occupied up to
high Landau levels (the filling factor is about 50 for the
experiments of Ref.~\onlinecite{man02}). We suppose that the
normalized eigenstates in the presence of disorder are
$\psi_\alpha({\bf r})$, with eigenvalues $\epsilon_\alpha$. Now we
turn on the microwave radiation. When an electron absorbs a
photon, it is excited by an energy $\omega$. In the absence of
disorder, conductivity is independent of the distribution of
electrons over Landau levels (Kohn's theorem \cite{koh61}) so
photo-excited electrons can make no additional contribution to the
dc current. In the presence of disorder, these electrons can be
scattered by impurities as they are excited. We use a
generalization of Fermi's golden rule to obtain a
position-dependent rate. To second order in the radiation, the
rate at which an electron is scattered from
initial state $\alpha'$ to final position $\bf r$ is
$w_{\alpha'}({\bf r})=2\pi \sum_\alpha |\psi_\alpha({\bf r})|^2
\delta(\epsilon_\alpha-\epsilon_{\alpha'}-\omega)\, |\langle
\alpha | eE\hat{x}|\alpha'\rangle|^2$,
where $\langle \alpha | eE\hat{x}|\alpha'\rangle$ is the matrix
element of the electric dipole operator
of the radiation field.  Initially, the
eigenstates are occupied with Fermi distribution function
$n_{F}(\epsilon_{\alpha'})$ about chemical potential $\mu$,
and we insert the factor
$[n_{F}(\epsilon_{\alpha'})-n_{F}(\epsilon_\alpha)]$
so that final states must be unoccupied. Then the additional current
density due to the radiation is
$\Delta J_{x}({\bf R}) = - e \int d\epsilon\, d^2\!\Delta
r\,[n_{F}(\epsilon)-n_{F}(\epsilon+\omega)]
\sum_{\alpha'} |\psi_{\alpha'}({\bf r}')|^2
\delta(\epsilon-\epsilon_{\alpha'})\,
w_{\alpha'}({\bf r})\,\Delta x$.
Here ${\bf R}=\frac{1}{2}({\bf r}+{\bf r}')$, $\Delta {\bf r}={\bf
r} - {\bf r}'=(\Delta x,\Delta y)$. For the average over disorder
(denoted by an overbar throughout this paper), we write this
expression as
\begin{eqnarray}
\Delta J_{x}({\bf R}) = -2\pi e \int d\epsilon\, d^2\!\Delta
r\,[n_{F}(\epsilon)-n_{F}(\epsilon+\omega)] && \nonumber \\
\times {\mathcal{N}}({\bf r}, \epsilon+\omega)
{\mathcal{N}}({\bf r}',\epsilon)M({\bf r}, {\bf r}')\,\Delta x &&
\label{eq:Mdef}
\end{eqnarray}
where ${\mathcal{N}}({\bf r}, \epsilon)=\overline{\sum_\alpha
|\psi_\alpha({\bf r})|^2\,\delta(\epsilon-\epsilon_\alpha)}$ is
the average local density of states, which in high Landau levels
obeys ${\mathcal{N}}(\epsilon + n\omega_c)={\mathcal{N}}(\epsilon)$.
$M({\bf r}, {\bf r}')>0$ is defined by extracting these density of
states factors, and is assumed to be independent of the energies
$\epsilon$, $\epsilon+\omega$, and to be a function of
$|\Delta{\bf r}|$ only. To measure the conductivity in linear
response, a small dc electric field, $E_{dc}$, is applied in the
$x$-direction. This tilts the energy spectrum as depicted in
Fig.~\ref{fig:toypicture}, and thus we assume that the only effect
on Eq.~(\ref{eq:Mdef}) is to make ${\mathcal{N}}({\bf r},
\epsilon)={\mathcal{N}}(\epsilon-eE_{dc}x)$. $\Delta J_x$ vanishes
to zeroth order in $E_{dc}$ by symmetry, and for linear response,
we expand to first order in $E_{dc}$, and divide by $E_{dc}$ to
find the longitudinal conductivity, $\sigma_{xx}$. We find that
the radiation-induced change in the longitudinal conductivity is
proportional to an integral of the partial derivative $(\partial
({\mathcal{N}}({\bf r},\epsilon+\omega){\mathcal{N}}({\bf
r}',\epsilon))/\partial \Delta x)_{\bf R}$. The density of states
can be roughly modelled by ${\mathcal{N}}(\epsilon) =
{\mathcal{N}}_{0} +
{\mathcal{N}}_{1}\cos(2\pi\epsilon/\omega_{c})$. The final
integral over $\epsilon$ can now be done, and at least for
$\omega/\omega_c$ large compared to $\mathcal{N}_{0}/\mathcal{N}_{1}$,
the result is
\begin{equation}
\Delta \sigma_{xx} \propto
-\sin(2\pi\omega/\omega_{c}),
\label{eq:toyRxx}
\end{equation}
with a positive coefficient. This form, which resembles the
derivative of the density of states $(\partial {\mathcal
N}/\partial\epsilon)|_{\epsilon=\omega}$, arises because the main
contribution is from initial states near the center of filled
broadened Landau levels, which are scattered to empty broadened
levels, and the available phase space is enhanced or diminished
for $\Delta x$ positive or negative, depending on the energy
change $\omega$ modulo $\omega_c$ (see Fig.~\ref{fig:toypicture}).
It is clear from experiment that $\sigma_{xy}$ is nearly 100
times larger than $\sigma_{xx}$ and is not significantly affected
by the radiation. Therefore, inverting the conductivity tensor
yields $\rho_{xx} \approx \rho_{xy}^{2} \sigma_{xx}$, which has
the period and phase of the oscillations observed in experiment.

While the above treatment is highly oversimplified, it indicates that
disorder plays a crucial role and may be all that is necessary to
obtain the radiation-induced oscillations.  This suggests that a
diagrammatic (Kubo formula) calculation of the conductivity,
including radiation and disorder but neglecting electron-electron
interactions, will be sufficient to reproduce the effect. This
calculation is presented below.

\begin{figure}
\centerline{\resizebox{3.375in}{!}{\includegraphics{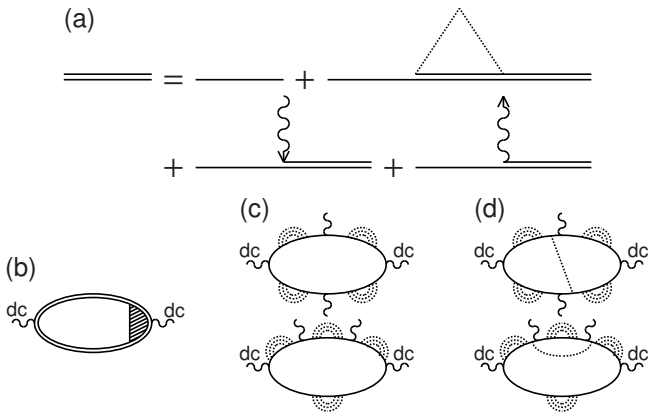}}}
\caption{Diagrams for (a) Green's function and (b) polarization bubble including
radiation and disorder within SCBA.  Disorder lines are dotted.
Photon lines are curvy.  The vertex of the polarization bubble is dressed with
ladders of disorder lines.  When the no-photons-crossed-by-disorder-lines conserving
approximation is made, diagrams like those in (c) are included while those in
(d) are neglected.}
\label{fig:diagrams}
\end{figure}

In the Landau gauge and the rotating-wave approximation (we neglect both
counter-rotating and guiding-center terms in the coupling to radiation),
the Hamiltonian is
\begin{eqnarray}
\lefteqn{H = \sum_{k,n} n\omega_{c}\, c_{nk}^{\dagger}c_{nk}
+ \sum_{k,n;k^{\prime},n^{\prime}} c_{nk}^{\dagger} c_{n^{\prime}k^{\prime}}
V_{n,k;n^{\prime},k^{\prime}}} \nonumber \\
&& + \frac{eE\ell}{\sqrt{2}} \sum_{k,n} \sqrt{n}
\left( c_{nk}^{\dagger}c_{n-1,k} e^{-i\omega t}
+ c_{n-1,k}^{\dagger}c_{nk} e^{i\omega t} \right)
\label{eq:Ham}
\end{eqnarray}
where $n$ is the Landau level index, $k$ is the $y$-component of
momentum, $c_{nk}$ is the electron annihilation operator,
$V_{n,k;n^{\prime},k^{\prime}}$ are the matrix elements of the
disorder potential $V_{\rm imp}({\bf r})$, and $E$ is the
magnitude of the electric field component of the microwave
radiation.  Following the calculation of Ando \cite{and74} for the
zero-radiation case, we shall include disorder within the
self-consistent Born approximation (SCBA) and assume
$\overline{V_{\rm imp}({\bf r})V_{\rm imp}({\bf r}^{\prime})} =
(2\gamma/m^{*}) \delta({\bf r}-{\bf r}^{\prime})$ where
$\gamma=1/2\tau$ is the elastic scattering rate in zero magnetic
field. This assumption of $\delta$-correlated disorder, while
somewhat inappropriate for the long-ranged impurity potentials
associated with modulation doping, significantly simplifies the
problem by eliminating the momentum dependence of the self-energy.
Yet, as we shall see, it manages to capture the important physics
rather well.  To all orders in the disorder and radiation, the
Green's functions are given by the diagrams in
Fig.~\ref{fig:diagrams}(a).  Since the radiation terms connect
neighboring Landau levels, evaluating these diagrams for the
retarded or advanced Green's functions,
$G^{R,A}_{nm}(t_{1},t_{2})$, yields a matrix equation in Landau
level indices $n$ and $m$.  (The Green's functions are
proportional to an identity matrix in the momentum label $k$,
which is therefore dropped.) The presence of radiation renders
this an inherently non-equilibrium problem which requires the
approach of Kadanoff and Baym and Keldysh (see
Refs.~\cite{hau96,kad62,ram86} for details).  This involves the
``lesser'' Green's function, $G^{<}_{nm}(t_{1},t_{2})$, which is
the only place where the distribution of electrons enters, and
which is evaluated from the same set of diagrams via a slightly
more complex procedure.  It is convenient to define
\begin{equation}
\bar{G}^{R,A,<}_{nm}(t,\mathcal{T}) = e^{in\omega t_{1}}
G^{R,A,<}_{nm}(t_{1},t_{2}) e^{-im\omega t_{2}}
\label{eq:Gbar}
\end{equation}
where $t \equiv t_{1}-t_{2}$, $\mathcal{T} \equiv (t_{1}+t_{2})/2$, and
$\bar{G}^{R,A,<}_{nm}(z,\mathcal{T})$, are the Fourier transforms in $t$.
After some manipulation, we obtain the following system of equations:
\begin{eqnarray}
\sum_{p} \biggl[ \left( z - n(\omega_{c}-\omega)
- \Sigma^{R,A}(z+n\omega) \pm i\delta \right) \delta_{np}
\;\;\;\;\;\;\;\;\;\; && \nonumber \\
- \frac{eE\ell}{\sqrt{2}}
\left( \sqrt{n} \,\delta_{n,p+1} + \sqrt{p} \,\delta_{n+1,p} \right)
\biggr] \bar{G}^{R,A}_{pm}(z) = \delta_{nm} &&
\label{eq:Gret}
\end{eqnarray}
\begin{equation}
\bar{G}^{<}_{nm}(z) \!=\! \sum_{p} \bar{G}^{R}_{np}(z)
\bigl[ n_{F}(z+p\omega)2i\delta
+ \Sigma^{<}(z+p\omega) \bigr] \bar{G}^{A}_{pm}(z)
\label{eq:Gless}
\end{equation}
\begin{equation}
\Sigma^{R,A,<}(z) = \frac{\gamma\omega_{c}}{\pi} \sum_{n}
\bar{G}^{R,A,<}_{nn}(z-n\omega)
\label{eq:Sigma}
\end{equation}
where $\delta \rightarrow 0^{+}$.
The Eqs.~(\ref{eq:Gret}-\ref{eq:Sigma}) have been displayed for the
steady state case, in which all Green's functions are independent of the
average time, $\mathcal{T}$.  The general $\mathcal{T}$-dependent situation
leads to more complicated expressions which can be obtained as discussed
in Ref.~\onlinecite{hau96}.
Given the Green's functions, the conductivity is obtained
via the Kubo formula,
$\sigma_{ij} = -\lim_{\Omega \rightarrow 0} \mbox{Im}\, \Pi_{ij}^{R}(\Omega)/\Omega$,
where $\Pi_{ij}^{R}$ is the retarded current-current correlation function,
and $i,j=\{x,y\}$.  Diagrammatically, this means evaluating the polarization
bubble in Fig.~\ref{fig:diagrams}(b) where the vertex is dressed by ladders of
impurity lines.

An exact solution is complicated by the self-consistent nature of
the above equations.  We shall simplify by neglecting vertex
corrections to the polarization bubble.  To do so within a
conserving approximation,
we must also neglect all diagrams in which impurity lines cross
photon insertions (see Fig.~\ref{fig:diagrams}(c,d)).  Note that
this is equivalent to replacing the fully self-consistent
$\Sigma$'s in the above equations with $\Sigma$'s calculated in
the absence of radiation but still self-consistent in the
disorder.  Thus, if we define $\Sigma(z)$ to be the self-energy
calculated by Ando \cite{and74} for the zero-radiation case, then
$\Sigma^{R}(z) \rightarrow \Sigma(z)$, $\Sigma^{A}(z) \rightarrow
\Sigma^{*}(z)$, and $\Sigma^{<}(z) \rightarrow -2i n_{F}(z)
\mbox{Im}\Sigma(z)$. This approximation also has the virtue of
mimicking energy loss mechanisms so that a steady state
$\mathcal{T}$-independent solution does indeed exist
\cite{approxfootnote}, and can be obtained directly from
Eqs.~(\ref{eq:Gret},\ref{eq:Gless}). The radiation enters only
when the Green's functions are calculated from the $\Sigma$'s. In
the absence of vertex corrections, it is straightforward to
calculate the conductivity from the Green's functions.  We find
that
\begin{eqnarray}
\left\{ \!\!\! \begin{array}{c} \sigma_{xx} \\ \sigma_{xy} \end{array} \!\!\! \right\}
= -\frac{e^{2}\omega_{c}^{2}}{4\pi^{2}} \sum_{n,m}
\sqrt{(n \!+\! 1)(m \!+\! 1)} \int \! dz \;\;\;\;\;\;\;\; && \nonumber \\
\times \biggl[ \mbox{Im}\bar{G}^{<}_{n+1,m+1}(z) \frac{d}{dz}
\left\{ \!\!\! \begin{array}{c} \mbox{Im} \\ \mbox{Re} \end{array} \!\!\! \right\}
\bar{G}^{R}_{nm}(z \!+\! \omega) \; && \nonumber \\
\pm\, \mbox{Im}\bar{G}^{<}_{nm}(z) \frac{d}{dz}
\left\{ \!\!\! \begin{array}{c} \mbox{Im} \\ \mbox{Re} \end{array} \!\!\! \right\}
\bar{G}^{R}_{n+1,m+1}(z \!-\! \omega) \biggr]
\label{eq:sigmaxxsigmaxy}
\end{eqnarray}
and obtain the resistivity by inverting the $\sigma$-matrix.

\begin{figure}
\centerline{\resizebox{3.5in}{!}{\includegraphics{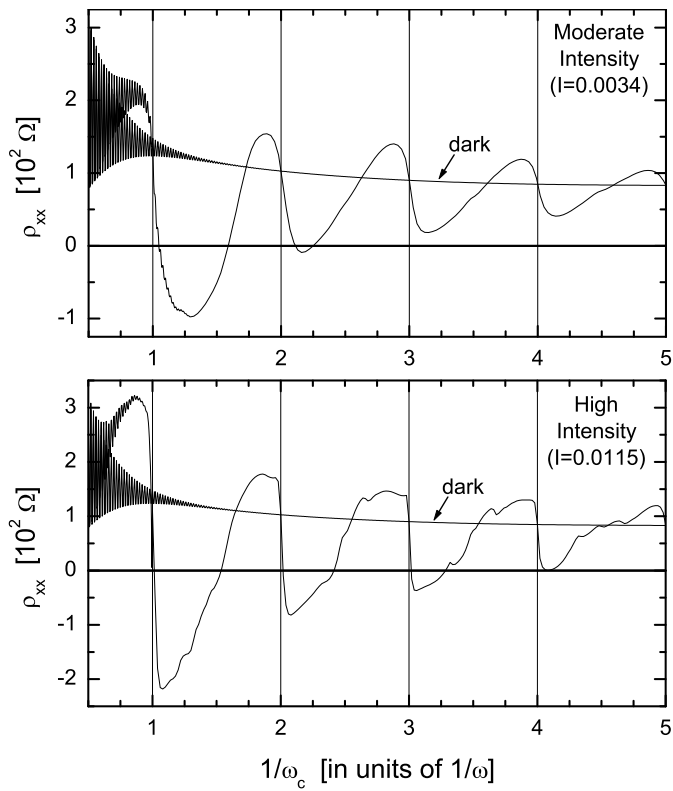}}}
\caption{Calculated radiation-induced resistivity oscillations.
We plot $\rho_{xx}$ vs $1/\omega_{c}$
at fixed $\omega$ for $\mu=50\omega$, $k_{B}T=\omega/4$, $\gamma=0.08\omega$,
and three values of radiation intensity (in units of $m^{*}\omega^{3}$):
$I=0$ (dark), $I=0.0034$ (upper panel), and $I=0.0115$ (lower panel).
For computational purposes, the energy spectrum
is cutoff at 20 Landau levels above and below the chemical potential.
The high-frequency oscillations seen at small $1/\omega_{c}$ are the
familiar SdH oscillations with period $1/\mu$.}
\label{fig:Rxx}
\end{figure}

Within the approximations discussed above, and using parameter
values appropriate to the experiments of Ref.~\onlinecite{man02},
we have calculated $\rho_{xx}$ numerically to all orders in
radiation and disorder.  Results are presented in
Fig.~\ref{fig:Rxx}, where we plot $\rho_{xx}$ versus
$1/\omega_{c}$ for fixed $\omega$.  We consider three
values of radiation intensity (power per unit area in units of
$m^{*}\omega^{3}$): $I=0$ (dark), $I=0.0034$ (moderate intensity),
and $I=0.0115$ (high intensity).  The dark resistivity exhibits
only the familiar SdH oscillations which have period $1/\mu$ and
decay away as $\omega_{c}$ becomes small compared with the
temperature.  For moderate intensity, we find a
pronounced radiation-induced oscillation of $\rho_{xx}$.
In agreement with experiment, the period of
oscillation is $1/\omega$ and minima are found {\it near}
$\omega/\omega_{c}=\mbox{integer}+1/4$. We note,
however, that the $1/4$ phase shift is not universal, varying
between $0$ and $1/2$ depending upon disorder and intensity. A
more robust feature is the presence of zeros of the oscillation at
the integer values of $\omega/\omega_{c}$. Unlike the SdH
oscillations, the radiation-induced oscillations are {\it not}
sensitive to the temperature-broadening of the electron
distribution about the Fermi level.  $T$-dependence
therefore derives from the the $T$-dependence of the
scattering, a feature absent from the present calculation due to
our neglect of interaction effects.  The magnitude of the
radiation-induced oscillations can easily exceed the dark
resistivity, leading to regions of negative total resistivity.
This is reasonable for a non-equilibrium system, and a similar
effect has been observed experimentally in semiconductor
superlattices \cite{kea95}, but it seems that some additional
physics is required for these negative-resistivity minima to
become the zero-resistance states observed in the present
experiments. For high intensity radiation, additional features
appear which have yet to be discovered experimentally. These
likely correspond to multi-photon processes by which an electron
absorbs $m$ photons and is promoted by $n$ Landau levels.

Since the simplifying approximations
discussed above, employed to make the calculation tractable,
are not well controlled, we cannot expect our results to
be quantitatively accurate.  However, the fact that even this simplified calculation
captures the qualitative form of the experimental results encourages
us that we have identified the essential physics of the problem.
Due to the assumption of $\delta$-correlated disorder, even the zero-radiation
(dark) resistivity, $\rho_{xx}^{0}$, does not
agree quantitatively with experiment.  As the calculation neglects the
difference between the transport lifetime, $\tau_{tr}$, and the single particle
lifetime, $\tau$, it overestimates $\rho_{xx}^{0}$ by a factor of
$\tau_{tr}/\tau$, equal to about 50 in the present case.  Since the dark
Hall resistivity, $\rho_{xy}^{0}$, is approximately independent of disorder in
the $B$-field regime of interest, the calculated value of $\rho_{xy}^{0}$
is correct, but effectively too small compared
to $\rho_{xx}^{0}$.  This quantitative error in the dark resistivity has two
important consequences.
When we calculate the radiation-induced change in the Hall resistivity,
$\Delta \rho_{xy}$, we find that it is also oscillatory and of comparable magnitude
to $\Delta \rho_{xx}$.  Since $\rho_{xy}^{0}$ is effectively too small, we find a
total Hall resistivity that exhibits noticeable radiation-induced oscillations.
For analogous reasons, while our calculations yield oscillations
in $\sigma_{xx}$ which grow linearly with radiation intensity, the intensity-dependence
is muted when we invert the conductivity matrix to obtain $\rho_{xx}$.
However, correcting for the missing factor of 50, it is clear that the
effect of radiation on $\rho_{xy}$ is negligible and that the oscillations in
$\rho_{xx}$ grow linearly with intensity, which is what was observed experimentally.

In conclusion, we understand the radiation-induced
magnetoresistance oscillations recently observed in the 2DEG
to be a consequence of photo-excited disorder-scattered electrons
contributing to the dc resistivity in an oscillatory manner inherited
from the broadened Landau level structure of the energy spectrum.
Our diagrammatic calculation yields a radiation-induced oscillation
of $\rho_{xx}$ that has the correct period and phase.
We find that for radiation of sufficient magnitude, the amplitude of the
oscillation can be made large enough that the total resistivity is
negative in the vicinity of the minima.  This effect is reminiscent
of the experimentally observed zero-resistance states but
indicates that additional physics, perhaps localization or
electron-electron interactions, is required to understand this
aspect of the observed phenomenon.  Our calculations at high intensity
reveal multi-photon effects, the observation of which should be a goal
for future experiments.  More advanced calculations, going beyond
the simplifying approximations employed herein and including the
interaction effects responsible for energy relaxation and
temperature dependence, are left for future work.

We are grateful to K. Lehnert who brought to our attention the experiments
of Ref.~\onlinecite{kea95}, which helped inspire the physical picture
discussed in this Letter.  We are also grateful to R. Willett for raising the
possibility of negative resistance in these systems.  We thank R. G. Mani,
R. Shankar, and R. Willett for helpful discussions.  This work was supported by
NSF Grants DMR-0103639 (A.C.D.), DMR-0098226 (S.S. and A.C.D.), and
DMR-0196503 (S.M.G. and A.C.D.).

After submitting this Letter for publication, we learned
that a related physical picture and a calculation of the
conductivity to second order in the radiation, with a
prediction of negative conductivity, were previously
reported by Ryzhii {\it et al.\/} \cite{ryz70,ryz86}.


\begin{thebibliography}{99}
\bibitem{zud01} M. A. Zudov {\it et al.\/}, Phys.\ Rev.\ B {\bf 64}, 201311 (2001)
\bibitem{man02} R. G. Mani {\it et al.\/}, Nature {\bf 420}, 646 (2002)
\bibitem{zud02} M. A. Zudov {\it et al.\/}, Phys.\ Rev.\ Lett.\ {\bf 90}, 46807 (2003)
\bibitem{koh61} W. Kohn, Phys.\ Rev.\ {\bf 123}, 1242 (1961)
\bibitem{and74} T. Ando, J.\ Phys.\ Soc.\ Japan {\bf 37}, 1233 (1974)
\bibitem{hau96} H. Haug and A. Jauho, {\it Quantum Kinetics in Transport
    and Optics of Semiconductors} (Springer, Berlin, 1996)
\bibitem{kad62} L. P. Kadanoff and G. Baym, {\it Quantum Statistical Mechanics}
    (Benjamin/Cummings, Massachusetts, 1962)
\bibitem{ram86} J. Rammer and H. Smith, Rev.\ Mod.\ Phys.\ {\bf 58}, 323 (1986)
\bibitem{approxfootnote} In the quantum Boltzmann equation, this
approximation corresponds to having a given state
repopulated by an ``in'' scattering rate from states at other
momenta and Landau levels with a radiation-independent thermal
occupation.
\bibitem{kea95} B. J. Keay {\it et al.\/}, Phys.\ Rev.\ Lett.\ {\bf 75}, 4102 (1995)
\bibitem{ryz70} V. I. Ryzhii, Fiz.\ Tverd.\ Tela {\bf 11}, 2577 (1969)
    [Sov.\ Phys.\ Solid State {\bf 11}, 2078 (1970)]
\bibitem{ryz86} V. I. Ryzhii {\it et al.\/}, Fiz.\ Tekh.\ Poluprovodn.\
    {\bf 20}, 2078 (1986) [Sov.\ Phys.\ Semicond.\ {\bf 20}, 1299 (1986)]
\end{thebibliography}
\end{document}